%% file: eprint.tex
\documentclass[10pt]{article}
\usepackage{graphicx}
\usepackage{amsmath}

\def\Title#1{\begin{center} {\Large #1 } \end{center}}
\def\Author#1{\begin{center}{ \sc #1} \end{center}}
\def\Address#1{\begin{center}{ \it #1} \end{center}}

\newcommand\pubblock{\rightline{\begin{tabular}{l} Proceedings of the Fifth Annual LHCP\\ \pubnumber\\
         \pubdate  \end{tabular}}}

\newenvironment{Abstract}{\begin{quotation} \begin{center} 
             \large ABSTRACT \end{center}\bigskip 
      \begin{center}\begin{large}}{\end{large}\end{center} \end{quotation}}

\newenvironment{Presented}{\begin{quotation} \begin{center} 
             PRESENTED AT\end{center}\bigskip 
      \begin{center}\begin{large}}{\end{large}\end{center} \end{quotation}}

\input econfmacros.tex

\usepackage{color}

\newcommand\pubnumber{ CMS CR-2017/358 } 

\newcommand\pubdate{\today}

\def\support{\footnote{on behalf of the CMS Collaboration}}

\begin{document}

\large
\begin{titlepage}
\pubblock

\vfill
\Title{Measurement of jet properties in CMS}
\vfill

\Author{Patrick~L.S.~Connor\support}
\Address{DESY, Notkestrasse 87, 22607 Hamburg, Germany}
\vfill
\begin{Abstract}
    We present measurements of the inclusive jet production at centre-of-mass energies of 8 and 13 TeV, and of multijets at 8 TeV.
    These measurements allow to constrain PDFs and the strong coupling constant.
    Two measurements of the azimuthal correlations at 8 and 13 TeV are also presented, testing higher order QCD calculations.
\end{Abstract}
\vfill

\begin{Presented}
The Fifth Annual Conference\\
on Large Hadron Collider Physics \\
Shanghai Jiao Tong University, Shanghai, China\\ 
May 15-20, 2017
\end{Presented}
\vfill
\end{titlepage}
\def\thefootnote{\fnsymbol{footnote}}
\setcounter{footnote}{0}

\normalsize 

\section{Introduction}

We summarise the six most recent jet measurements in CMS \cite{CMSdet} at 8 and 13 TeV.
Analog measurements at 8 and 13 TeV are presented together.

\section{Inclusive jet analyses at 8 and 13 TeV}

The double differential cross section as a function of the transverse momentum and the rapidity is given by
\begin{align}
    \frac{\mathrm{d}^2 \sigma}{\mathrm{d} p_T \mathrm{d} y} &= \frac{1}{\epsilon \mathcal{L}_\text{int}^\text{eff}} \frac{N_\text{jets}}{\Delta p_T \Delta |y|}
\end{align}

At 8 TeV \cite{Khachatryan:2016mlc}, the luminosity for the high $p_T$ region is $19.7~\text{fb}^{-1}$, while for the low $p_T$ region with dedicated low-pile-up runs, it is $5.6~\text{pb}^{-1}$.
The measurement is performed for large cone size jets only with the anti-$k_T$ algorithm \cite{Cacciari:2008gp}.
The rapidity coverage is $0 < |y| < 4.7$.

A similar measurement was performed with the first data at 13 TeV \cite{Khachatryan:2016wdh}.
The luminosity is $71~\text{pb}^{-1}$ in the central region ($|y| < 3.0$) and $44~\text{pb}^{-1}$ in the forward region ($3.2 < |y| < 4.7$).
The rapidity coverage and binning are kept the same as for the analysis at 8 TeV. 
The measurement at 13 TeV is performed for two cone size radii: $R=0.4$ and $R=0.7$.

The measurements are compared to predictions from MC event generators and from fixed-order parton-level calculations.
In general, the uncertainty related to the correction of the jet energy scale is of the order of a percent.

\begin{figure}
    \centering
    \includegraphics[width=0.5\textwidth]{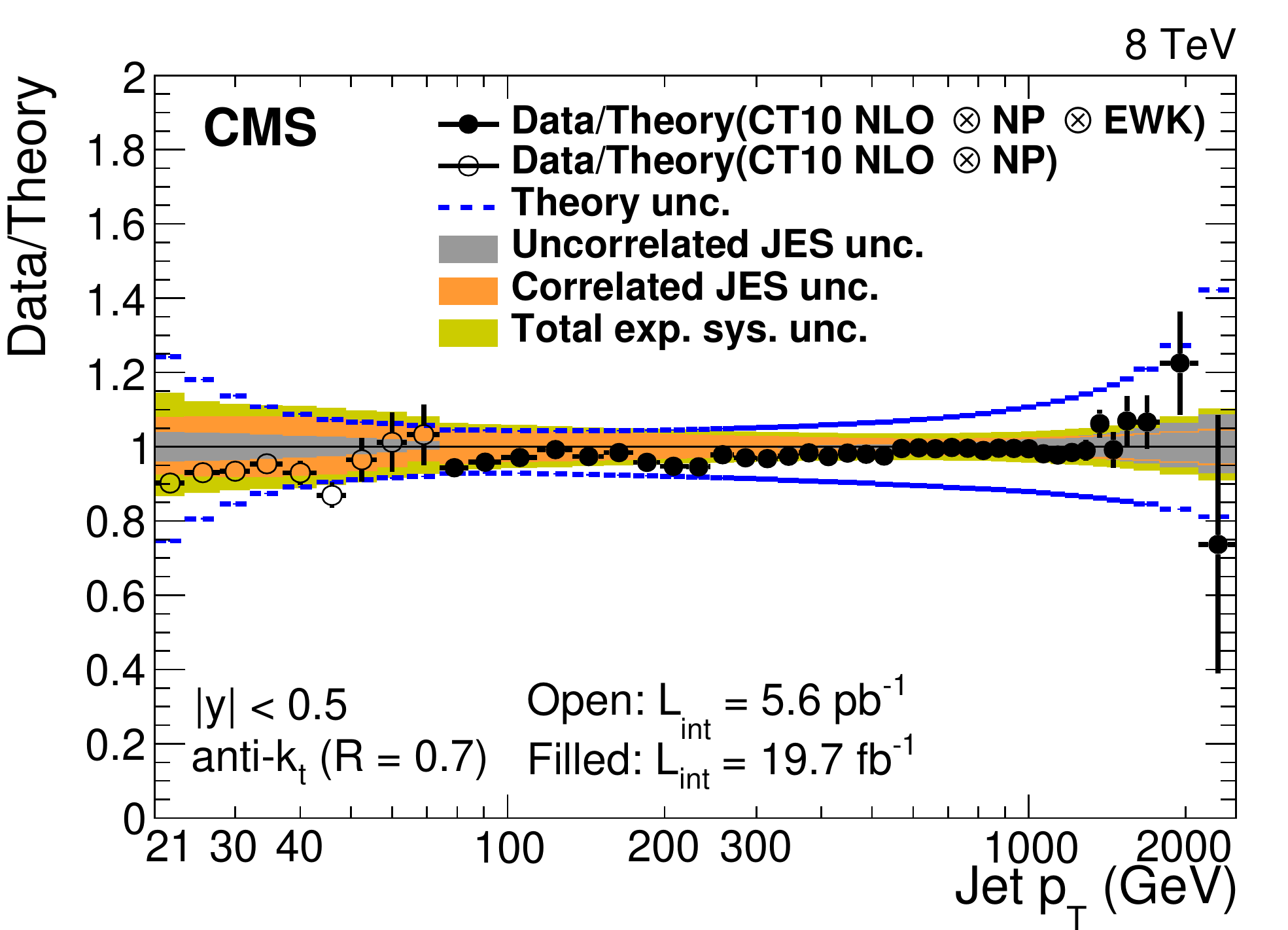}~
    \includegraphics[width=0.4\textwidth]{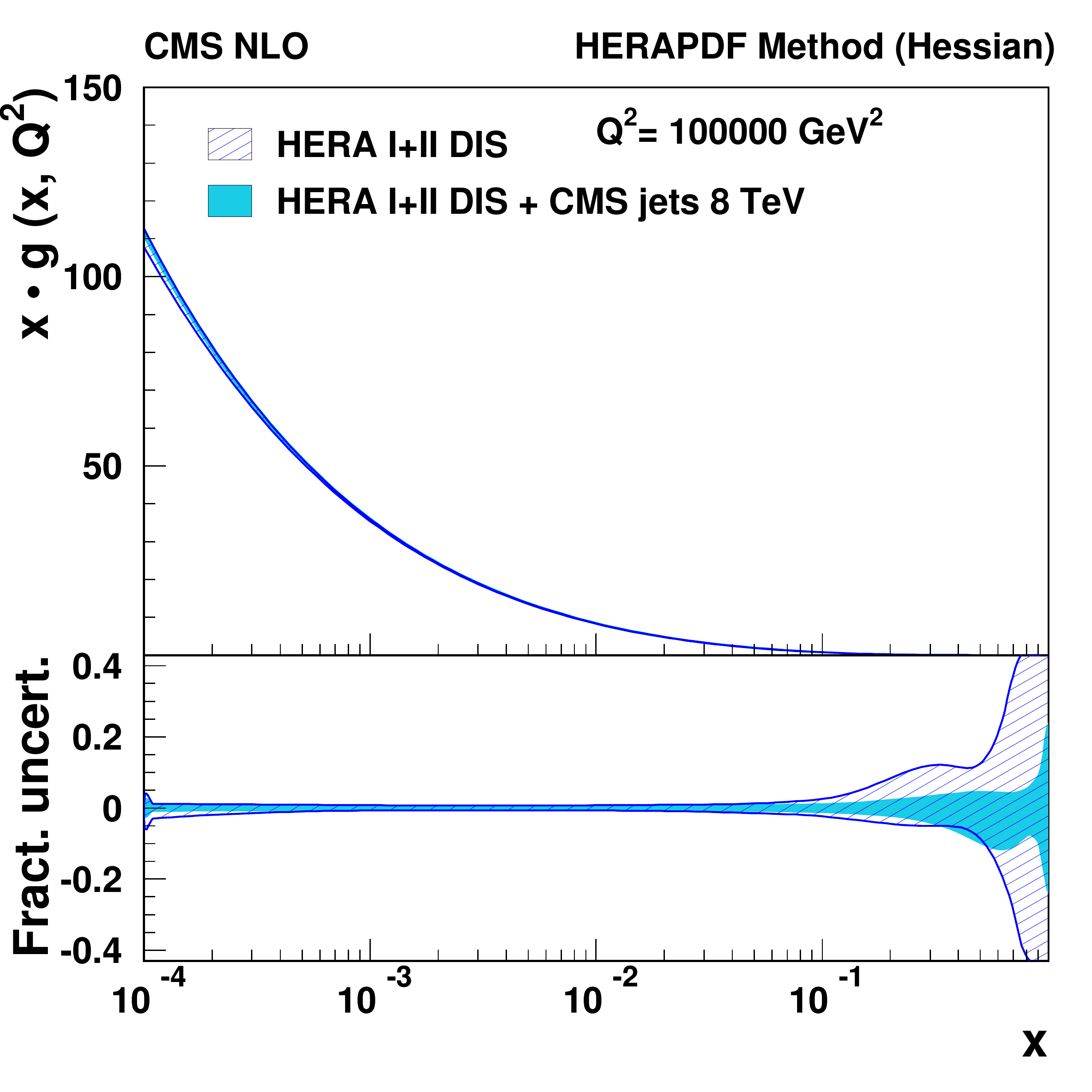}

    \caption{Left: comparison to fixed-order parton-level calculations of the inclusive jet production at $8~\text{TeV}$. Right: gluon PDF before and after inclusion of inclusive jet data \cite{Khachatryan:2016mlc}.}
    \label{fig:ratio_NLO_calculation_8TeV}
\end{figure}

The fixed-order calculation includes the electroweak and non-perturbative QCD corrections.
We observe a very good agreement over the two orders of magnitude.
The gluon PDF can be constrained with the inclusive jet data, and $\alpha_S$ can be extracted together with the fit of the PDF, as an additional parameter, with a value of $0.1164^{+0.0093}_{-0.0073}$.
The measurement and the PDF fit are illustrated in Fig.~\ref{fig:ratio_NLO_calculation_8TeV}.

The comparison to fixed-order parton-level calculations (not shown here) agrees better for a large cone size radius than for a small cone size radius.
This is understood as being related to missing higher order calculations. 

The comparison of the measurement with predictions from \textsc{PowHeg} + P8 (NLO + PS) shows very good agreement for both cone sizes at 13 TeV.

\section{Multijet analysis at 8 TeV}

\begin{figure}
    \centering
    \includegraphics[width=0.45\textwidth]{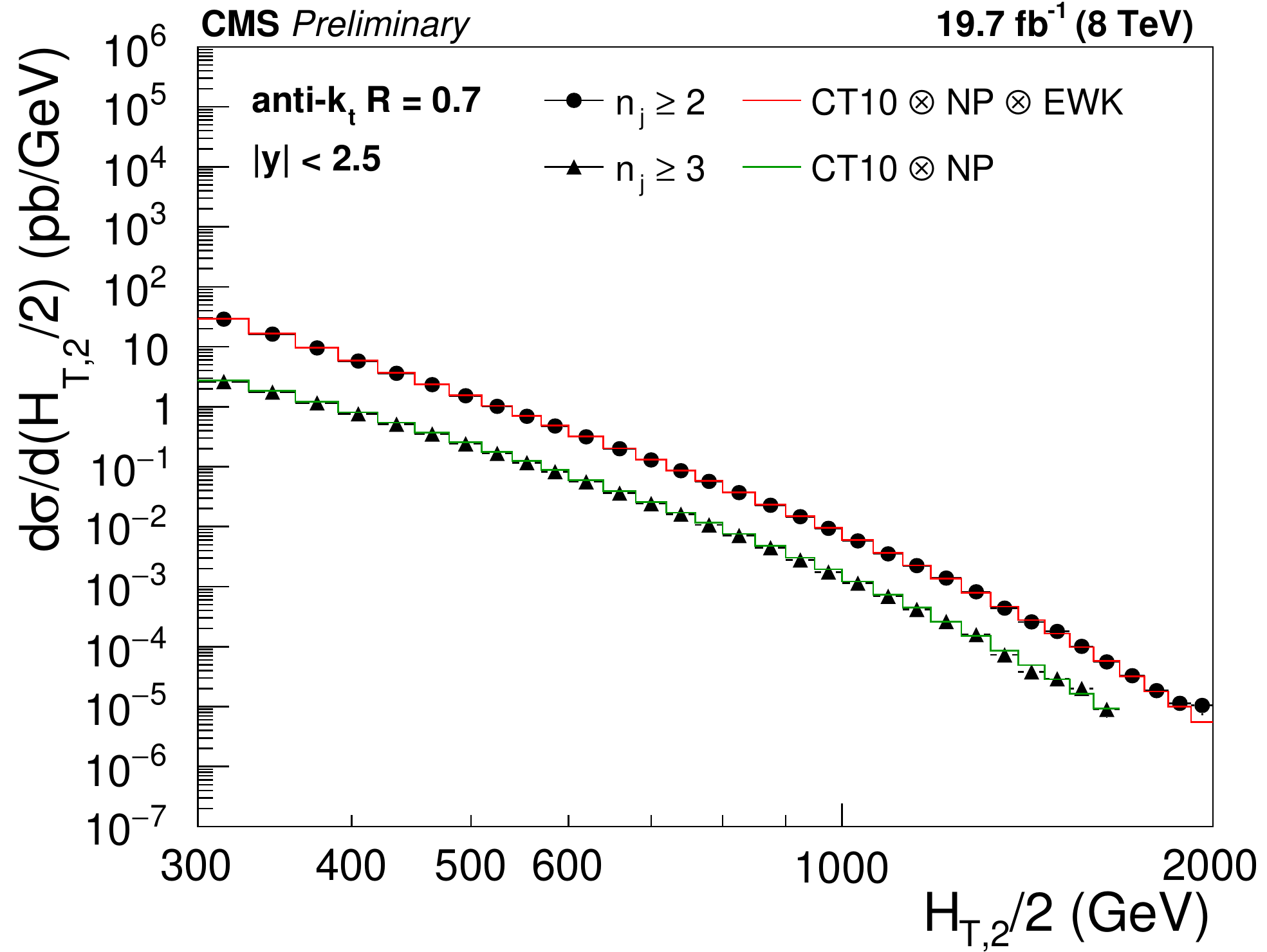}~
    \includegraphics[width=0.45\textwidth]{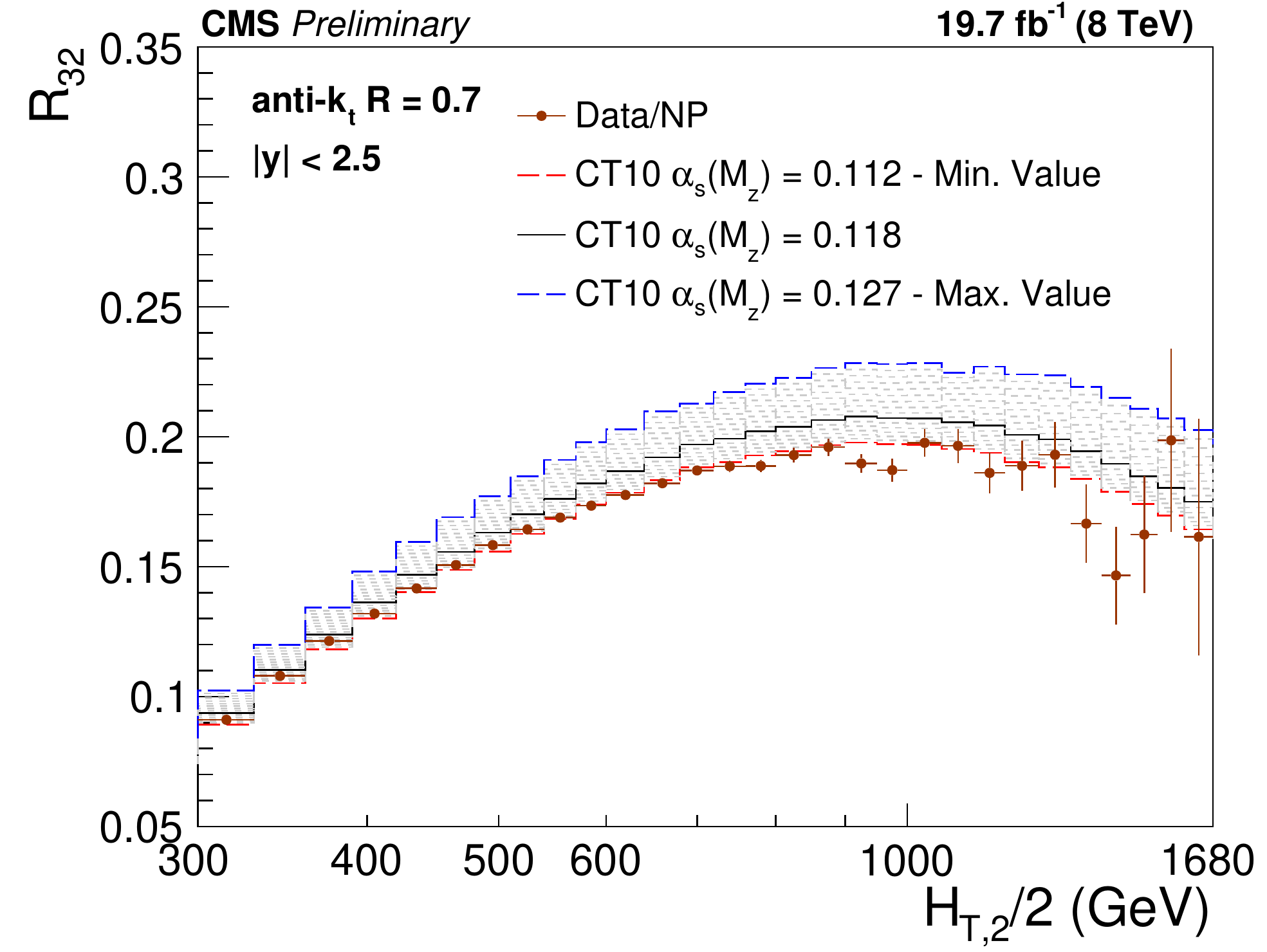}

    \caption{Measurement of the di-jet and tri-jet production at $8~\text{TeV}$ \cite{CMS:2017tvp}.}
    \label{fig:CMS_multijet}
\end{figure}

The measurement of the multijet production was performed as a function of $H_{T,n} = \sum^n_{i=1} p_{T,i}$ with all the jets of an event in the considered phase space \cite{CMS:2017tvp}:
\begin{align}
    \frac{\mathrm{d} \sigma}{\mathrm{d} (H_{T,2}/2)} &= \frac{1}{\epsilon \mathcal{L}_\text{int}^\text{eff}} \frac{N_\text{events}}{\Delta (H_{T,2}/2)}
\end{align}
Since $R_{mn} = \frac{\sigma_{m-\text{jet}}}{\sigma_{n-\text{jet}}} \propto \alpha_S^{m-n} $, $\alpha_S$ can be extracted from the ratio of 3 and 2-jet measurements.
Some uncertainties cancel in the ratio. 
The selection consists of jets of $p_T > 150 ~\text{GeV}$ in $|y|<2.5$, clustered with the anti-$k_T$ algorithm with $R=0.7$.
Events where the leading jet would be in the forward region are vetoed.
The measurement is shown in Fig.~\ref{fig:CMS_multijet} and $\alpha_S=0.1150^{+0.0088}_{-0.0038}$ is extracted.

\section{Dijet triple differential cross section at 8 TeV}

The triple differential cross section of the di-jet production at $8~\text{TeV}$ consists in the following measurement \cite{CMS:2016wpr}:
\begin{align}
    \frac{\mathrm{d}^3 \sigma}{\mathrm{d} p_{T,\text{avg}} \mathrm{d} y^* \mathrm{d} y_b} &= \frac{1}{\epsilon \mathcal{L}_\text{int}^\text{eff}} \frac{N_\text{di-jet events}}{\Delta p_{T,\text{avg}} \Delta y^* \Delta y_b}
\end{align}
where
\begin{itemize}
    \item   $p_{T,\text{avg}} = \frac{1}{2} ( p_{T,1} + p_{T,2} )$ is the average transverse momentum of the di-jet system;
    \item   $y_b = \frac{1}{2}|y_1 + y_2|$ is the \emph{rapidity boost} of the di-jet system;
    \item   $y^* = \frac{1}{2}|y_1 - y_2|$ is the \emph{rapidity separation} of the di-jet system.
\end{itemize}
Anti-$k_T$ jets with $R=0.7$ are selected with $p_{T,\text{jet}} > 50 ~\text{GeV}$, $|y_\text{jet}| < 3.0$ and $p_{T,\text{avg}} > 133 ~\text{GeV}$.

The different regions of the phase space are then exploited to extract the strong coupling and to constrain PDFs: the central region (small $y_b$ and small $y^*$) is most suited for $\alpha_S$ extraction at high energy scales; in the boosted region (higher $y_b$ but small $y^*$), the high-$x$ region of PDFs can be better constrained; finally, in the region of large rapidity separation (small $y_b$ but higher $y^*$), PDF and detector effects can be better disentangled.

\begin{figure}
    \centering

    \includegraphics[width=0.49\textwidth]{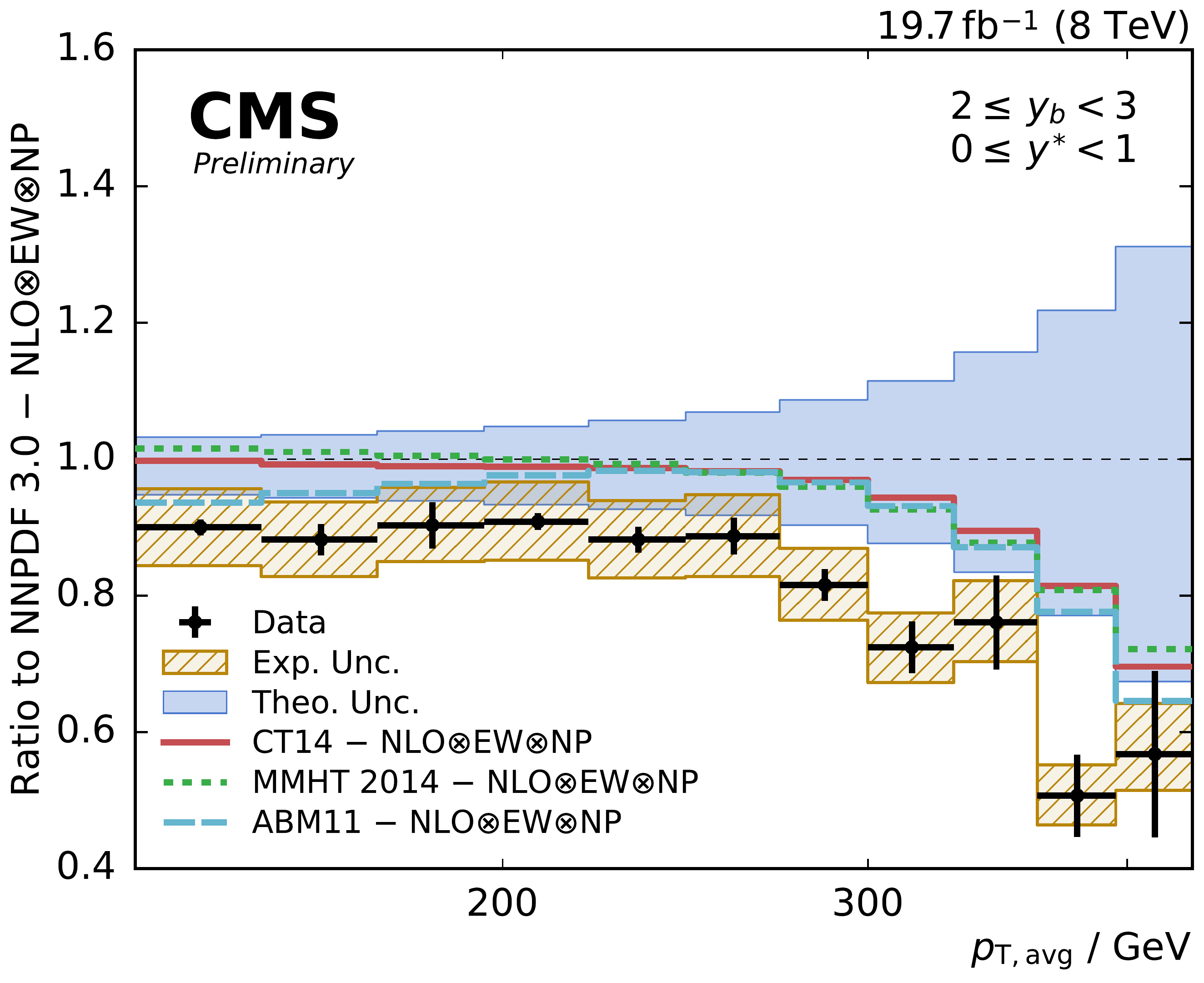}
    \includegraphics[width=0.49\textwidth]{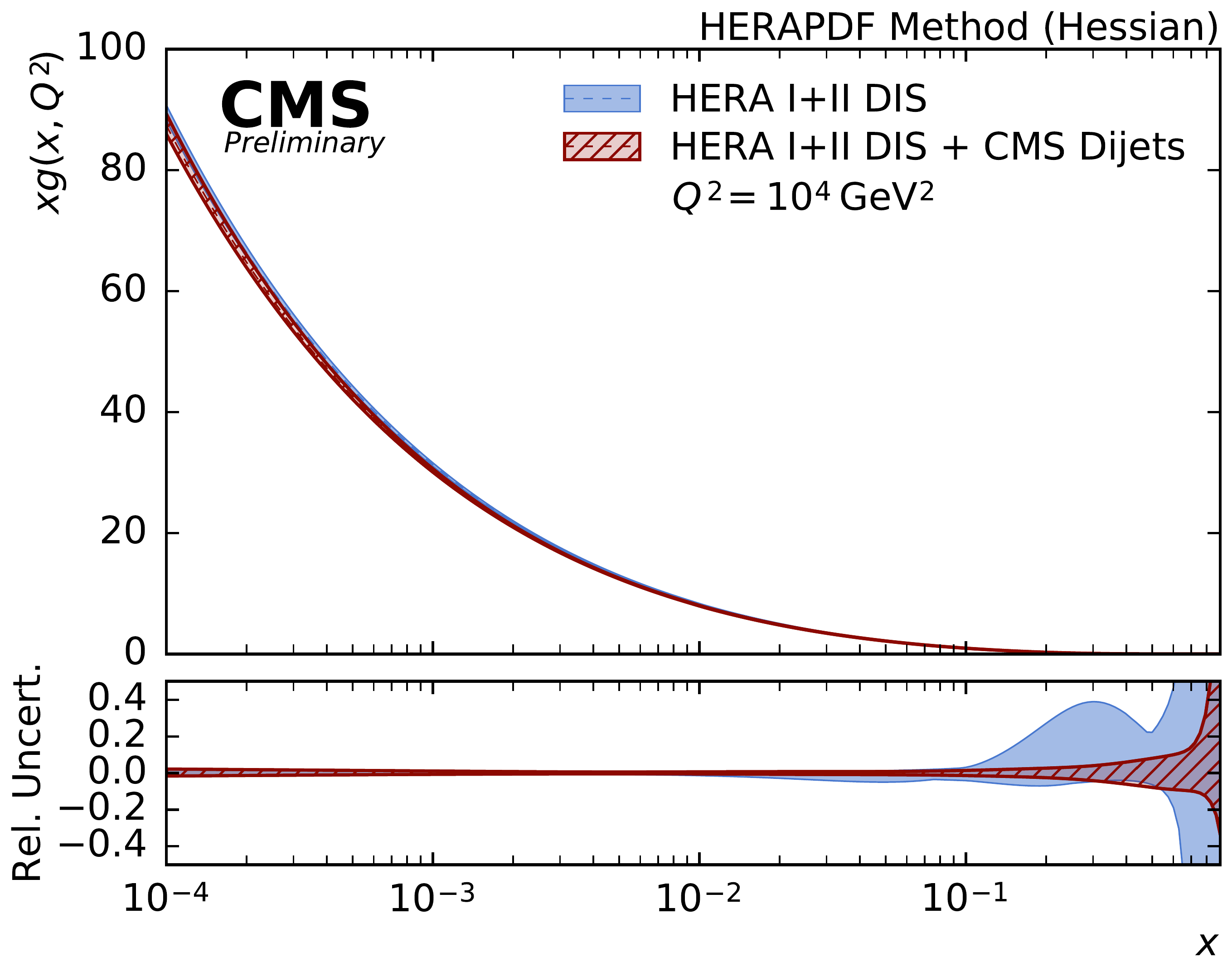}

    \caption{Left: di-jet triple differential cross section in the boosted region. Right: comparisons of the gluon PDF before and after inclusion of di-jet data \cite{CMS:2016wpr}.}
    \label{fig:CMS_3_diff_xsect_NLO_calc}
\end{figure}

The measurement is compared first to NLO parton-level calculations as well as predictions from MC event generators (\textsc{Herwig}~7 and \textsc{Pythia}~8).
In general, predictions are slightly overestimated at high transverse momentum in the boosted region.
Eventually, the di-jet data may be used to extract gluon PDFs.
This is illustrated in Fig.~\ref{fig:CMS_3_diff_xsect_NLO_calc}.

\section{Azimuthal correlations at 8 and 13 TeV}

The azimuthal correlations allow to investigate higher-order QCD corrections: 
the more extra radiations, the less correlated the two leading jets.

CMS has published two measurements:
\begin{enumerate}
    \item   the measurement of the azimuthal correlations between the two leading jets at 8 TeV with a jet reconstructed with the anti-$k_T$ algorithm of $R=0.7$ \cite{Khachatryan:2016hkr};
    \item   the analog measurement at 13 TeV for a radius of 0.4 \cite{CMS:2017cqo}; in addition, the same measurement is also performed with minimum jet multiplicities of 3 and 4 jets in order to be more sensitive to higher order effects; finally, correlations between the subleading and the closest third- or fourth-leading jet are measured, also in order to increase the sensitivity to parton showers.
\end{enumerate}

\begin{figure}
    \includegraphics[width=0.49\textwidth]{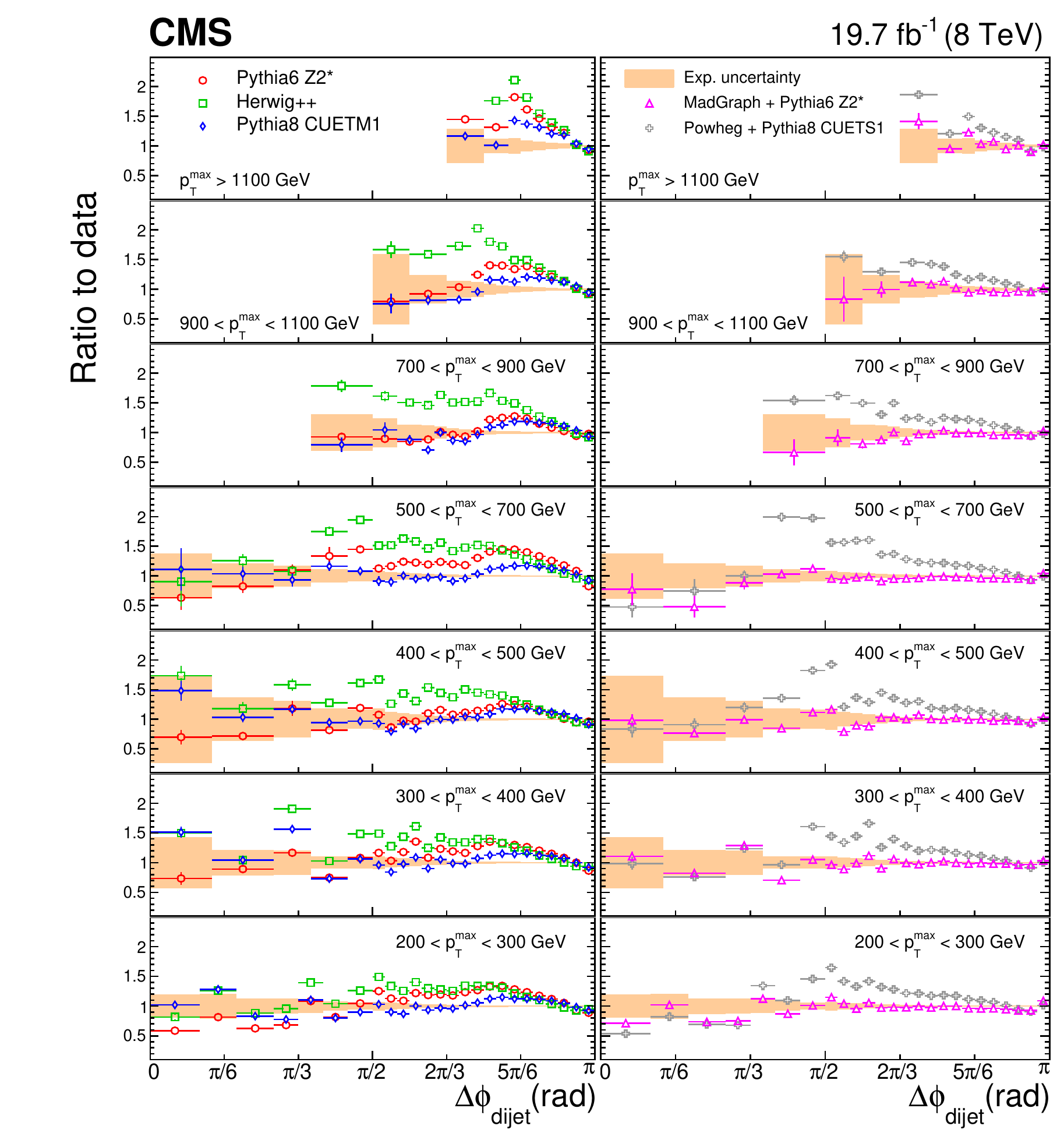}
    \includegraphics[width=0.45\textwidth]{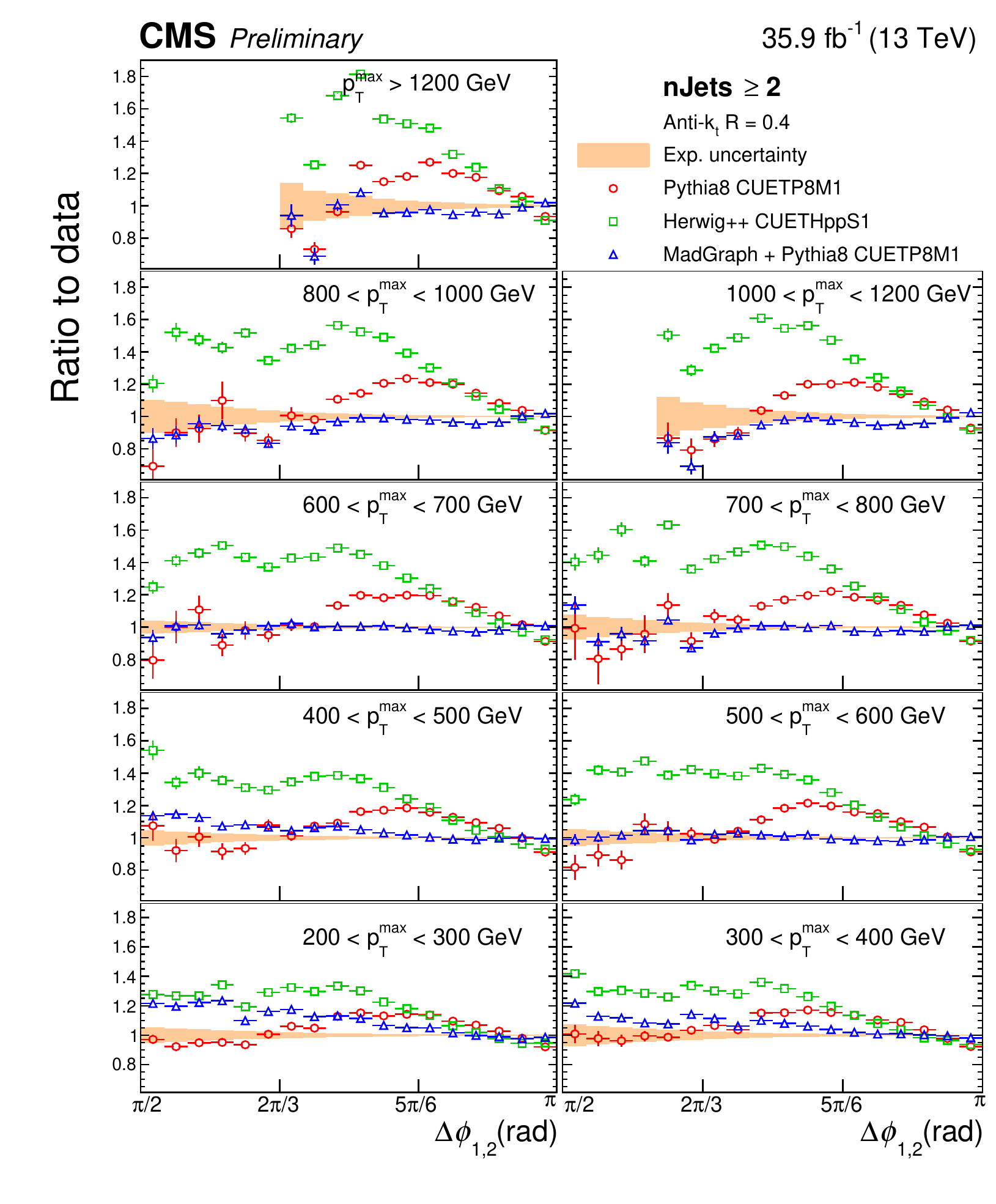}

    \caption{Analog measurements at 8 and 13 TeV (with respective cone size radii) for the azimuthal correlations \cite{CMS:2017cqo}.}
    \label{fig:azimuthal_correlations}
\end{figure}

Here, given the large amount of results reported in these analyses, we only show corresponding measurements at 8 and 13 TeV in Fig.~\ref{fig:azimuthal_correlations}.

\section{Conclusions}

In general, the QCD predictions describe well the measurements.
With jet data, gluon PDFs can be significantly improved at high $x$.
The value of $\alpha_S$ is also extracted.
The azimuthal correlations in multijet events illustrate the importance of higher-order QCD corrections.

\end{document}

%% file: econfmacros.tex
\def\beq{\begin{equation}}
\def\eeq#1{\label{#1}\end{equation}}
\def\eeqn{\end{equation}}

\def\beqa{\begin{eqnarray}}
\def\eeqa#1{\label{#1}\end{eqnarray}}
\def\eeqan{\end{eqnarray}}

\let\bar=\overbar

\def\Dslash{\not{\hbox{\kern-4pt $D$}}}
\def\dslash{\not{\hbox{\kern-2pt $\del$}}}

\def\msb{{\bar{\ssstyle M \kern -1pt S}}}

